\documentclass[lettersize,journal]{IEEEtran}
\usepackage{amsmath,amsfonts}
\usepackage{array}
\usepackage{tikz}
\usepackage{amsmath,mathtools}
\usepackage{amssymb}
\usepackage{enumitem}
\usepackage{booktabs}
\usepackage{graphicx}
\usepackage{amsmath}
\usepackage[table]{xcolor}
\usepackage{algorithm}
\usepackage{subfig}
\usetikzlibrary{shadows, positioning, arrows.meta, calc, chains, fit, backgrounds, fadings}
\usepackage{booktabs}
 \usepackage{tabularx}
 \usepackage{array}

\usepackage{algpseudocode}
\usepackage{boxedminipage}
\usepackage{fancyhdr}
\usetikzlibrary{arrows.meta, positioning, calc, shadows, backgrounds, fit}
\usepackage{multirow}
\usetikzlibrary{arrows.meta, chains, shapes.symbols, shadows}

\usepackage{mathptmx}
\usepackage{array}
\usepackage{textcomp}
\usetikzlibrary{positioning}
\usetikzlibrary{arrows.meta}
\usetikzlibrary{calc}
\usetikzlibrary{shapes.geometric}
\usetikzlibrary{fit}
\usepackage{stfloats}
\usepackage{csquotes} 
\usepackage{url}
\usepackage{verbatim}
\usepackage{cite}
\usepackage{eqnarray}
\usepackage{stfloats}
\usepackage{url}
\usepackage{verbatim}
\usepackage{graphicx}
\def\BibTeX{{\rm B\kern-.05em{\sc i\kern-.025em b}\kern-.08em
    T\kern-.1667em\lower.7ex\hbox{E}\kern-.125emX}}
\usepackage{balance}
\usepackage{pifont}

\begin{document}
\title{The Network That Thinks: Kraken* and the Dawn of Cognitive 6G}

\author{
    Ian F. Akyildiz, \textit{Life Fellow, IEEE} and Tuğçe Bilen, \textit{Member, IEEE}
    \thanks{Ian F. Akyildiz is with Truva Inc., Alpharetta, GA 30022, USA (e-mail: ian@truvainc.com).}%
    \thanks{Tuğçe Bilen is with the Department of Artificial Intelligence and Data Engineering, Istanbul Technical University, Istanbul, Turkey (e-mail: bilent@itu.edu.tr).}%
    \thanks{*KRAKEN (Knowledge-centric, Reasoning, And goal-oriented Knowledge NetworK): Inspired by the mythological sea creature, Kraken symbolizes a vast, interconnected intelligence capable of multi-layered reach and collective coordination across the 6G landscape.}
    \thanks{A more comprehensive version of this work providing detailed architectural analysis and deployment considerations has been submitted to Proceedings of IEEE.}
}


\maketitle
\begin{abstract}
Future sixth-generation (6G) networks must evolve beyond high-speed data delivery to support intelligent, context-aware services. Emerging applications such as autonomous transportation, immersive extended reality, and large-scale sensing require networks capable of interpreting context, anticipating system dynamics, and coordinating resources according to application objectives rather than relying solely on packet-level metrics. This article introduces \emph{Kraken}, a knowledge-centric architectural vision for enabling collective intelligence in 6G networks. Kraken integrates three complementary capabilities: semantic communication, which prioritizes the transmission of task-relevant information; generative reasoning, which enables predictive modeling of network and application dynamics; and goal-oriented optimization, which aligns resource allocation with application-level outcomes. These capabilities are organized within a three-plane architecture consisting of an Infrastructure Plane, an Agent Plane, and a Knowledge Plane. Together, these planes enable distributed network entities to perceive context, reason about future states, and coordinate actions through shared semantic representations. The architecture leverages emerging technologies such as O-RAN, network digital twins, and scalable MLOps pipelines, providing a practical evolutionary path from current 5G systems toward knowledge-centric 6G infrastructures. Three representative scenarios illustrate how Kraken improves efficiency and responsiveness in autonomous mobility, immersive XR services, and infrastructure monitoring. The article also outlines key research challenges and discusses the transition from today’s data-centric networks toward knowledge-centric collective intelligence in future 6G systems.
\end{abstract}

\begin{IEEEkeywords}
6G, network architecture, semantic communication, generative AI, goal-oriented networks, collective intelligence.
\end{IEEEkeywords}

\thispagestyle{fancy}

\pagestyle{fancy}
\fancyhf{}
\fancyhead[C]{\normalsize Submitted to IEEE Communications Magazine}
\renewcommand{\headrulewidth}{0pt}

\section{Why 5G's Success Creates 6G's Challenge}

The success of 5G is undeniable. Within only a few years, it has enabled multi-gigabit wireless access, dense device connectivity, and low-latency communication at large scale \cite{Akyildiz2022_XR}. These advances have supported services once considered long-term ambitions, including wireless industrial automation, immersive media delivery, and early forms of connected autonomy. Yet this progress also reveals a deeper limitation. As communication performance improves, the central bottleneck is no longer solely how fast networks can move bits, but whether they can deliver the information that actually matters for a task at the moment it is required.

This shift becomes evident in emerging 6G scenarios. Consider autonomous vehicles approaching an occluded intersection. Safe coordination does not require transmitting every raw sensor sample to a remote server; instead, it depends on identifying the observations essential for coordination and delivering them with minimal delay. A similar challenge arises in industrial augmented reality, where a worker wearing a headset expects digital guidance to respond instantly to motion and context, and even small delays disrupt the experience. Infrastructure monitoring systems exhibit the same pattern: distributed sensors generate large volumes of data, while only a small fraction is relevant for real-time anomaly detection. Across these examples, the implication is clear: future network performance will be judged less by the amount of transported data and more by whether the network enables effective decisions \cite{9955525, 10597087}.

In such environments, a purely data-centric design begins to show its limits. Traditional networks assume that once bits are reliably delivered, higher layers determine their value. This model works well for applications such as file transfer, web browsing, and media streaming, but becomes far less effective when service success depends on real-time inference, coordination, and control. In these settings, not all information is equally important, and decisions cannot always wait for complete data delivery.

\begin{figure*}[h]
    \centering
    \includegraphics[width=0.7\linewidth]{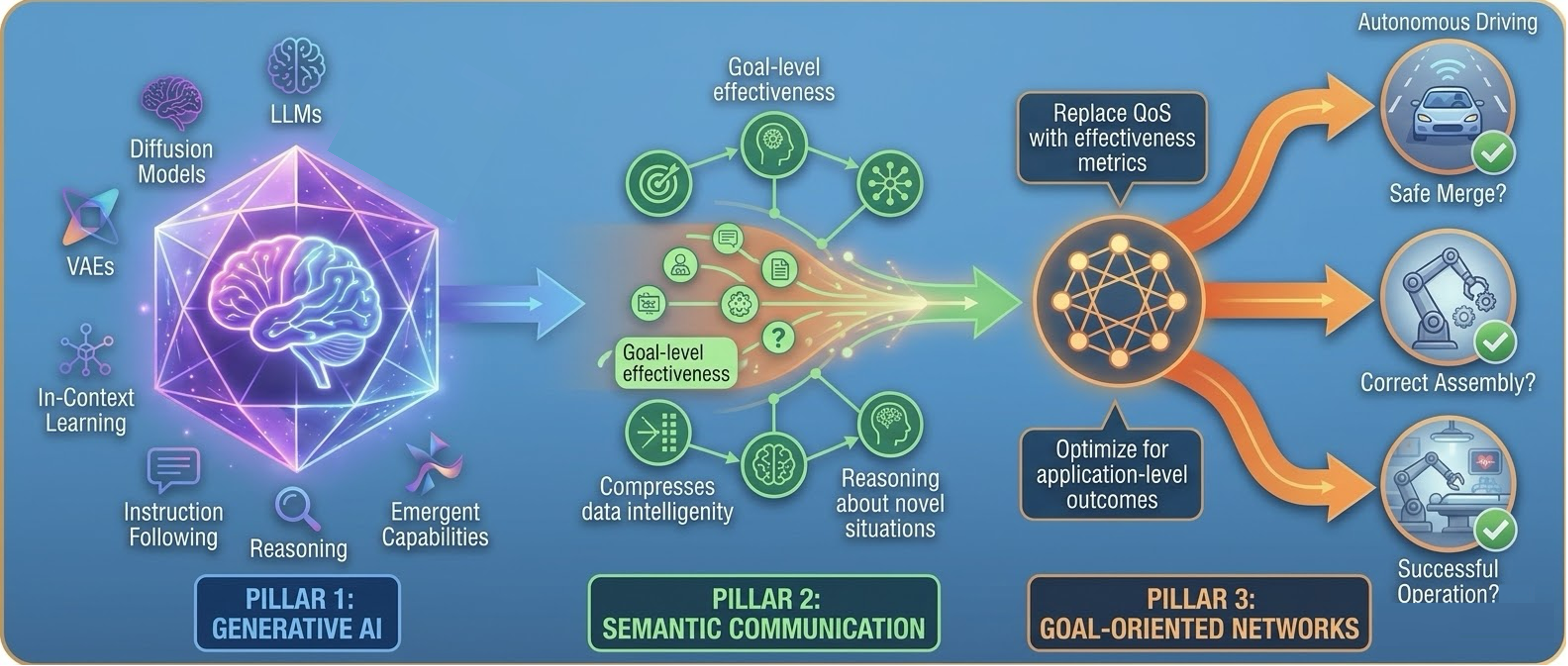}
    \caption{From data delivery to task effectiveness: generative AI provides predictive reasoning and world-model capabilities, semantic communication filters and compresses task-relevant information, and goal-oriented networking aligns network behavior with application-level outcomes. Their joint operation enables the transition from passive communication infrastructure to knowledge-centric collective intelligence.}
    \label{fig:capabilities}
\end{figure*}

Kraken is motivated by this shift. Rather than treating communication, intelligence, and control as separate functions, it integrates three tightly connected capabilities. The first is semantic communication, which prioritizes the transmission of task-relevant information rather than raw data volume. The second is generative reasoning, which enables network entities to build predictive models and evaluate possible future states. The third is goal-oriented networking, which evaluates network behavior according to application-level outcomes rather than conventional QoS indicators. As illustrated in Fig.~\ref{fig:capabilities}, these capabilities become most powerful when operating together as a unified system.

The key message is not simply that future networks require more artificial intelligence or stronger data compression. The deeper requirement is coordinated intelligence: the ability to reason about what matters, communicate relevant information efficiently, and act to improve task success. In this sense, 6G is not only a faster generation of wireless networks; it represents a transition toward systems that incorporate knowledge, prediction, and intent as native design elements.

Kraken is proposed as an architectural response to this challenge (a detailed architectural analysis and representative use cases are provided in \cite{akyildiz2026kraken}). It builds on practical enablers such as Open RAN and network digital twins while extending them toward a broader knowledge-centric framework in which communication, reasoning, and coordination operate together. This perspective provides a realistic path from today's 5G infrastructure toward future 6G systems that are not only highly connected but also cognitively capable.

\section{Toward Knowledge-Centric Networking}

The transition discussed in the previous section reflects a broader shift in networking research. As emerging applications increasingly rely on real-time inference, coordination, and control, several research directions have sought to move beyond purely data-centric communication. These include semantic communication, AI-driven network management, goal-oriented networking, multi-agent control, knowledge-based networking, and foundation models for telecommunications. While each direction provides valuable capabilities, they have largely evolved independently rather than forming a unified architectural framework. Kraken integrates these strands into a knowledge-centric architecture that enables distributed collective intelligence in future 6G systems, as summarized in Table~I.

\subsection{Semantic Communication}

Traditional communication theory focuses on reliable symbol transmission over noisy channels. However, many emerging applications require transmitting only the information relevant to a downstream task rather than reconstructing the original signal exactly. This observation has motivated semantic communication, where systems are optimized according to task effectiveness instead of bit-level fidelity. Recent work has explored neural joint source–channel coding approaches such as DeepJSCC, showing that learned communication systems can preserve perceptual or semantic content under bandwidth constraints \cite{8723589}. Broader studies have further formalized task-oriented communication as a framework for optimizing transmitted representations according to downstream inference or control objectives \cite{9955525}.

Despite these advances, most existing work focuses on point-to-point communication and single-task scenarios. Mechanisms for exchanging semantic information across distributed agents and aligning interpretations among heterogeneous network entities remain largely unexplored. Kraken extends semantic communication beyond link-level optimization by embedding semantic representations into a distributed architecture that enables coordinated knowledge exchange.

\subsection{AI and Machine Learning for Network Management}

Artificial intelligence has become a key tool for managing modern communication networks. Machine learning techniques have been widely applied to traffic prediction, anomaly detection, mobility management, and resource allocation. In particular, deep reinforcement learning has demonstrated significant improvements over rule-based heuristics in dynamic environments \cite{8714026}, while large-scale learning approaches increasingly support network analytics and operational automation \cite{10867389}. However, most existing approaches rely on discriminative models that map observations directly to predictions or actions, often lacking explicit representations of network dynamics.

Kraken introduces generative reasoning into network control through distributed Generative Network Agents that maintain predictive world models of network behavior. These models allow agents to anticipate potential future states and synthesize coordination strategies rather than reacting solely to observed conditions.

\begin{table*}[t]
\centering
\caption{Comparison of Kraken with representative intelligent networking paradigms highlighting the integration of semantic communication, generative reasoning, and goal-oriented coordination.}
\label{tab:comparison}
\scriptsize
\renewcommand{\arraystretch}{1}
\setlength{\tabcolsep}{7pt}

\begin{tabular}{p{1.6cm} p{2.6cm} p{2.6cm} p{2.6cm} p{2.6cm} p{3.2cm}}

\rowcolor{gray!30}
\textbf{Dimension} 
& \textbf{Semantic Communication}
& \textbf{AI/ML Management}
& \textbf{Goal-Oriented Networking}
& \textbf{Multi-Agent Networking}
& \textbf{Kraken Architecture} \\ 
\midrule

\rowcolor{gray!10}
Primary Focus
& Task-relevant transmission
& Data-driven network optimization
& Application-level objectives
& Distributed decision-making
& Knowledge-centric collective intelligence \\ 

Information Representation
& Compressed semantic features
& Network metrics / KPIs
& Task-level performance indicators
& State vectors
& Knowledge objects (facts, intents, uncertainty) \\ 
\hline

\rowcolor{gray!10}
Coordination Mechanism
& Point-to-point communication
& Centralized analytics
& Intent-driven optimization
& Reinforcement learning signals
& Knowledge exchange and world-model alignment \\ 

Decision Logic
& Task-aware encoding
& Predictive optimization
& Goal-driven control policies
& Distributed policy learning
& Generative reasoning with goal alignment \\ 
\hline

\rowcolor{gray!10}
Architectural Scope
& Link-level optimization
& Network management modules
& Service-layer optimization
& Distributed agents
& Unified three-plane architecture \\ 

\bottomrule

\end{tabular}
\end{table*}

\subsection{Goal-Oriented Networking}

Another emerging direction focuses on optimizing communication systems according to application-level goals rather than intermediate network metrics. In goal- or task-oriented networking, communication and control mechanisms are designed to maximize outcomes such as inference accuracy, control stability, or task success probability. This paradigm has gained increasing attention in 6G research, where communication systems are increasingly evaluated according to task effectiveness rather than raw throughput or latency \cite{10644029}.

However, most existing work primarily defines new optimization objectives without specifying how such goals can be implemented within large-scale distributed architectures. Kraken addresses this limitation by introducing a Knowledge Plane that translates service intents into coordination constraints for distributed agents.

\subsection{Multi-Agent Networking and Distributed Control}

Distributed decision-making has also been explored through multi-agent reinforcement learning and decentralized control frameworks. Applications include distributed spectrum sharing, interference coordination, and edge computing orchestration. Multi-agent learning enables network entities to adapt based on local observations while maintaining network-wide efficiency \cite{Oliehoek2016}. At the architectural level, initiatives such as O-RAN introduce programmable control loops through near-real-time and non-real-time RAN intelligent controllers \cite{doi:https://doi.org/10.1002/9781119847083.ch4}.

Nevertheless, most existing systems exchange measurements or engineered features rather than structured semantic knowledge. Kraken advances this paradigm by enabling distributed agents to exchange structured knowledge objects containing contextual information, uncertainty estimates, and intent attributes.

\subsection{Knowledge and Foundation Models in Networking}

Knowledge representations have also been explored through knowledge-graph-based network management and semantic networking approaches. Knowledge graphs support tasks such as root cause analysis, configuration validation, and anomaly detection in complex networks \cite{WANG2024124679}. More recently, foundation models have attracted attention in telecommunications research. Architectures such as NetGPT investigate the use of large language models for network diagnostics and automated management \cite{10466747}.

Kraken positions foundation models as epistemic priors within the Knowledge Plane while maintaining a constrained, verifiable control hierarchy, thereby enabling structured reasoning within operational network control loops.

\subsection{Positioning of Kraken}

Taken together, existing research has advanced several aspects of intelligent networking, including semantic communication, AI-driven management, distributed decision-making, and knowledge representations. However, these directions typically address isolated layers of the network architecture.

Kraken unifies these strands within a coherent framework for knowledge-centric 6G systems. Its three-plane architecture integrates semantic-aware infrastructure, distributed Generative Network Agents, and a global Knowledge Plane. Within this framework, semantic communication determines what information should be transmitted, generative reasoning enables predictive understanding of network dynamics, and goal-oriented coordination aligns network behavior with application objectives. This integration provides a concrete architectural foundation for distributed collective intelligence in future 6G networks.

\section{The Knowledge-Centric Shift: Three Capabilities That Matter}

Building on the research directions discussed in the previous section, a clear architectural shift is emerging. Future 6G networks will support services that require tight coupling among communication, sensing, computation, and control. In such environments, network performance cannot be evaluated solely through conventional metrics such as throughput or latency. Instead, communication systems must increasingly support application-level objectives, where the usefulness of transmitted information and the timeliness of network decisions directly influence service outcomes. This shift motivates a transition from traditional data-centric networking toward a knowledge-centric perspective that jointly considers communication, reasoning, and control.

Insights from artificial intelligence, control theory, and communication systems indicate that three capabilities are central to enabling this transition: semantic communication, generative reasoning through predictive world models, and goal-oriented network optimization. As illustrated in Fig.~\ref{fig:capabilities}, these capabilities enable networks to move beyond passive data transport toward coordinated decision-making across distributed entities.

\subsection{Transmitting Meaning Rather Than Raw Data}

Conventional communication systems emphasize reliable transmission and accurate signal reconstruction, treating packets as uniform data units regardless of contextual importance. This approach becomes inefficient when network resources are limited and the value of information depends on its relevance to a specific task.

Semantic communication addresses this limitation by incorporating task relevance directly into the communication process. Rather than optimizing only for signal reconstruction fidelity, the objective becomes preserving the information required for successful task execution. In practice, systems may exchange compact semantic representations, such as predicted trajectories or scene descriptors, instead of raw sensor or video streams. By prioritizing task-relevant information, semantic communication improves spectral efficiency while maintaining robustness under varying channel conditions.

\subsection{Predictive Reasoning Through Generative World Models}

While semantic communication determines what information should be transmitted, networks must also anticipate how conditions evolve over time. Mobility patterns, traffic demand, and channel dynamics strongly influence network performance, and purely reactive control strategies often struggle to maintain stable service quality.

Generative reasoning enables predictive decision-making by allowing networks to learn internal models of system dynamics. These world models capture relationships among network states, user behavior, and environmental factors, enabling evaluation of possible future scenarios and preparation of appropriate responses. Such predictive capabilities support proactive resource management, for example, anticipating user mobility, forecasting rendering demands in immersive applications, or detecting abnormal behavior in sensing infrastructures. As a result, network control evolves from reactive adaptation toward proactive coordination.

\subsection{Optimizing for Application-Level Objectives}

Even with meaning-aware communication and predictive reasoning, network operation requires clearly defined objectives. Traditional wireless systems typically optimize intermediate metrics such as throughput, latency, and packet loss, which do not always capture the ultimate goals of emerging applications. In many 6G scenarios, desired outcomes are defined directly at the application level. Autonomous mobility emphasizes safe coordination, extended reality requires perceptual stability and low motion-to-photon latency, and sensing infrastructures prioritize reliable anomaly detection.

Goal-oriented networking incorporates these objectives directly into the control loop. Resource allocation, scheduling, and routing decisions are therefore evaluated based on their contribution to task success rather than solely on protocol-level metrics. Measures such as semantic distortion, intent satisfaction, and task completion probability become part of the optimization framework, aligning network behavior with service outcomes.

Together, semantic communication, generative reasoning, and goal-oriented optimization form the conceptual foundation of the Kraken architecture. The following section describes how these principles are realized within the Kraken architectural framework.

\begin{figure*}[h]
\centering
\includegraphics[width=0.6\textwidth]{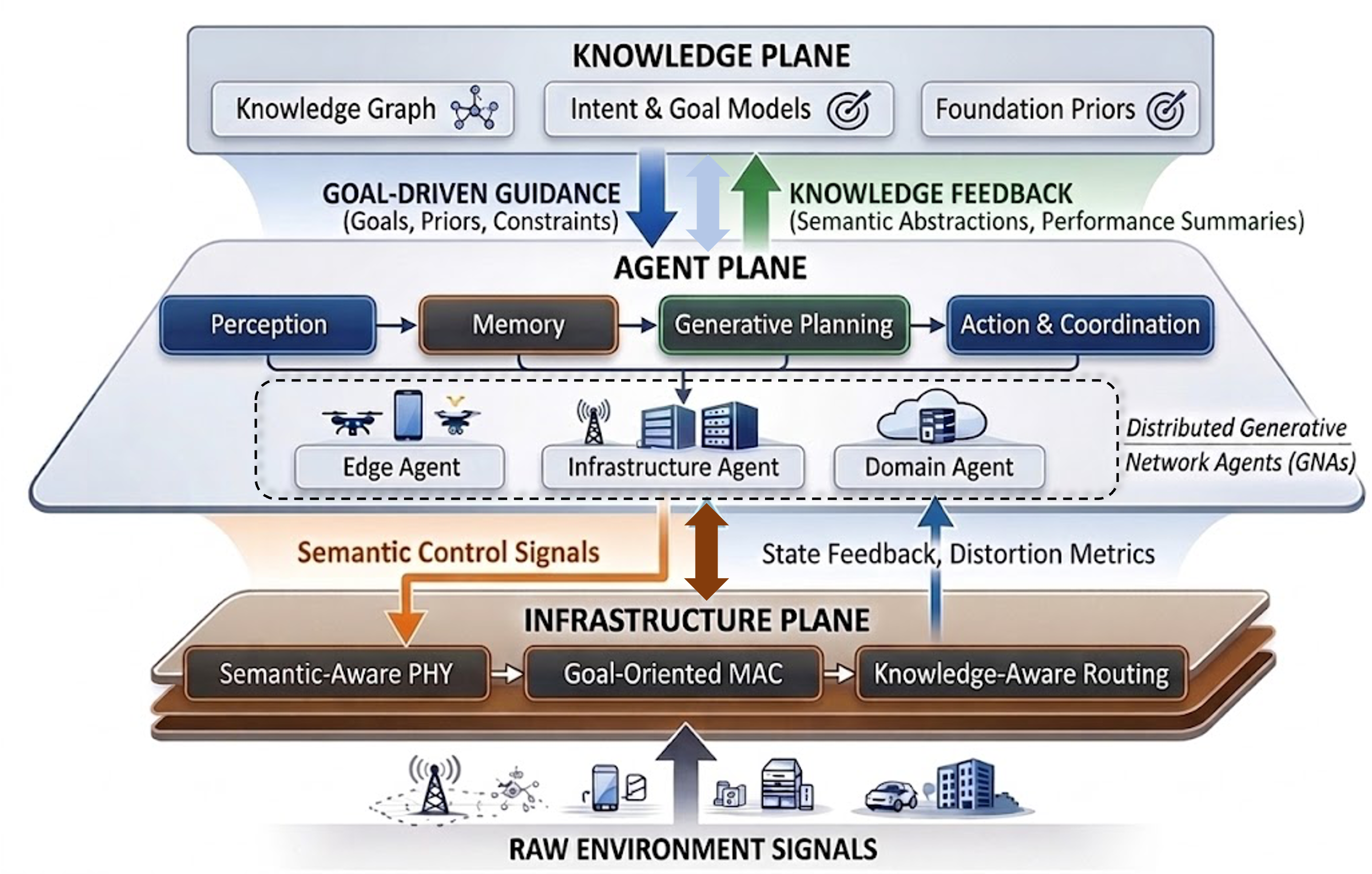}
\caption{Kraken's three-plane architecture. The Infrastructure Plane provides semantic-aware PHY/MAC/NET functions. The Agent Plane hosts Generative Network Agents with perception-memory-planning-action cycles. The Knowledge Plane maintains shared semantics, foundation model priors, and intent governance. Arrows show closed-loop information flow.}
\label{fig:architecture}
\end{figure*}

\section{Kraken: A Three-Plane Architecture for Collective Intelligence}

The three capabilities discussed in the previous section become operational through the Kraken architecture. Rather than concentrating intelligence within a single control entity, Kraken distributes cognition across three tightly coupled planes: the Infrastructure Plane, the Agent Plane, and the Knowledge Plane. Each plane performs a distinct role while interacting through closed control loops. Fig.~\ref{fig:architecture} illustrates how these layers transform raw environmental signals into coordinated, goal-aligned network actions.

\subsection{Infrastructure Plane: A Semantically-Aware Foundation}

The Infrastructure Plane extends conventional communication mechanisms with semantic awareness while remaining compatible with emerging 6G radio technologies. Physical-layer transmission, MAC scheduling, and routing decisions are augmented with information about task relevance and application intent.

At the physical layer, semantic contribution scores influence transmission parameters such as subcarrier allocation, coding redundancy, and transmission power. Information critical to downstream decisions, such as vehicle maneuver intents or XR gaze updates, can therefore receive stronger protection, while less relevant data is delivered with lighter overhead. At the MAC layer, scheduling considers semantic priority, task deadlines, and goal-contribution metrics alongside conventional channel conditions. Resources are therefore allocated to reduce semantic distortion rather than simply maximizing throughput. Retransmission policies also become task-aware; losing a pedestrian trajectory estimate, for example, is far more critical than losing background visual data. At the network layer, routing decisions incorporate task identifiers, latency sensitivity, and reliability requirements in addition to traditional metrics, favoring paths that better preserve semantic fidelity or reduce task-level distortion.

Importantly, the Infrastructure Plane executes control signals originating from higher layers without interpreting their semantic meaning. This separation preserves compatibility with existing protocol implementations while enabling semantic-aware communication behavior.

\subsection{Agent Plane: Distributed Network Cognition}

The Agent Plane introduces distributed intelligence through Generative Network Agents (GNAs). These agents are embedded across user equipment, base stations, edge servers, and core infrastructure, allowing cognition to emerge throughout the network rather than from a centralized controller.

Each GNA operates through a four-stage cognitive cycle. \textit{Perception} converts raw infrastructure measurements into structured semantic representations that describe inferred system states, contextual relationships, and confidence levels. For example, an autonomous vehicle agent interprets sensor observations as occupancy maps with predicted trajectories rather than raw signals. \textit{Memory} integrates these representations into a persistent knowledge structure, embedding observations within a distributed knowledge graph that captures historical context and enables information sharing among agents. \textit{Planning} relies on a generative world model to simulate possible future system states and evaluate alternatives such as mobility shifts, interference patterns, or traffic variations. Finally, \textit{Action} translates planning outcomes into coordinated control decisions executed through the infrastructure layers.

To ensure scalability, GNAs are deployed hierarchically. Edge-level agents within devices perform rapid local reasoning, infrastructure-level agents at base stations maintain awareness across multiple cells, and domain-level agents in regional or cloud infrastructure synthesize longer-horizon strategies and cross-domain coordination policies.

\subsection{Knowledge Plane: Global Semantic Coherence}

While agents reason locally, coherent system behavior requires a shared understanding of context and objectives. The Knowledge Plane provides this global coordination layer.

First, a \textit{distributed knowledge graph} defines shared ontologies, entity relationships, and semantic schemas across communication, mobility, and service domains, ensuring consistent interpretation of exchanged knowledge objects. Second, \textit{foundation models} capture statistical priors derived from large-scale network telemetry, including traffic dynamics, mobility patterns, and interference correlations, providing contextual guidance for agent planning while remaining outside strict real-time control loops. Third, an \textit{intent repository} translates operator-defined objectives into formal constraints and utility functions, allowing service-level agreements, safety policies, and operational requirements to shape agent decisions.

Through continuous monitoring of semantic distortion, task success rates, and fairness indicators, the Knowledge Plane performs meta-level feedback. When misalignment appears, it adjusts constraint parameters or refines semantic schemas to maintain consistent system-wide behavior across distributed agents.

\section{From Architecture to Deployment}

Architectural concepts alone cannot transform future networks. For a knowledge-centric architecture such as Kraken to become operational, it must be supported by practical deployment platforms, safe experimentation environments, and scalable model lifecycle management mechanisms. Three technological enablers play a central role in translating the architectural principles into deployable systems: Open RAN as the execution platform for distributed intelligence, Network Digital Twins as the validation environment, and MLOps pipelines for lifecycle governance of generative models.

\subsection{O-RAN as the Execution Platform}

Open RAN provides a natural operational environment for Kraken’s distributed intelligence. Its disaggregated architecture separates radio functions from control intelligence and enables programmable network management through standardized interfaces.

Within this ecosystem, infrastructure-level Generative Network Agents can operate as xApps hosted by the Near-Real-Time RAN Intelligent Controller (Near-RT RIC). These agents perform coordination functions on timescales ranging from tens of milliseconds to seconds, enabling adaptive scheduling, load balancing across neighboring cells, and interference mitigation under dynamic network conditions.

Longer-horizon reasoning tasks are handled by domain-level agents deployed as rApps within the Non-Real-Time RIC. These components manage network slicing policies, refine service intents, and coordinate resource allocation strategies across network segments. Standardized interfaces such as A1 and E2 connect these layers and support interoperability across multi-vendor deployments. In this architecture, Kraken does not replace the existing RAN framework but introduces a cognitive coordination layer operating above standardized control interfaces.

\subsection{Network Digital Twins for Safe Experimentation}

Deploying generative coordination mechanisms directly in operational networks introduces risks, since new policies may produce unintended interactions or instability. Network Digital Twins address this challenge by providing high-fidelity virtual replicas of the physical infrastructure.

These replicas mirror base station configurations, traffic dynamics, mobility patterns, and radio propagation conditions. Within this environment, Generative Network Agents can evaluate coordination strategies prior to deployment while enforcing safety constraints, stability requirements, and service-level objectives.

Rare operational events, such as cascading congestion or large-scale mobility surges, can also be synthesized to improve model robustness. Telemetry collected from operational networks continuously updates the twin, ensuring synchronization with real-world conditions and enabling iterative policy refinement before live deployment.

\subsection{MLOps for Scalable Model Lifecycle Management}

Knowledge-centric networking introduces a new operational challenge: managing large numbers of generative models distributed across heterogeneous network nodes. When thousands of agents operate across base stations, edge servers, and devices, manual configuration becomes infeasible.

Machine Learning Operations (MLOps) frameworks provide automated mechanisms for training, validation, deployment, and monitoring of these models. Distributed data collection pipelines enable learning from geographically diverse environments while preserving privacy through federated learning. Candidate models are validated within the Network Digital Twin before deployment to ensure compliance with operational constraints.

After deployment, monitoring mechanisms track model behavior and detect distributional drift caused by changes in traffic patterns, mobility dynamics, or environmental conditions. When degradation is detected, retraining or rollback procedures can be triggered automatically. Model compression techniques such as quantization, pruning, and knowledge distillation further allow generative capabilities to operate within the limited compute and energy budgets of edge devices.

\subsection{Alignment with 3GPP Standardization Efforts}

Although Kraken is presented as a conceptual architecture, many of its principles align with ongoing standardization activities within the 3GPP ecosystem and related industry initiatives. At the service level, the 3GPP Service and System Aspects group (SA1) has begun investigating intelligent services requiring tight integration of communication, sensing, and computing capabilities. These discussions emphasize context awareness, intent-driven service management, and support for emerging applications such as immersive media, connected automation, and autonomous mobility \cite{3gpp22_261}.

Within system architecture, the SA2 working group has introduced analytics-driven management functions such as the Network Data Analytics Function (NWDAF), which provides real-time and predictive analytics to network functions and represents an early step toward AI-assisted network operation \cite{3gpp23_288}. At the radio access level, the O-RAN Alliance architecture enables programmable RAN control through the RAN Intelligent Controller (RIC), supporting both near-real-time and non-real-time applications via xApps and rApps \cite{polese2023understanding}. This framework provides a practical execution environment for distributed intelligence components such as the Generative Network Agents envisioned in Kraken.

From this perspective, Kraken can be viewed as a forward-looking architectural abstraction that extends these developments. Semantic communication complements context-aware networking, generative agents extend analytics-driven intelligence, and the Knowledge Plane aligns with emerging efforts toward shared network intelligence and intent-based management. Together, these mechanisms transform Kraken from a conceptual vision into a practical framework capable of operating across large-scale 6G infrastructures.

\section{Kraken in Action: Three Representative 6G Scenarios}

\begin{figure}[h]
    \centering
    \includegraphics[width=0.95\linewidth]{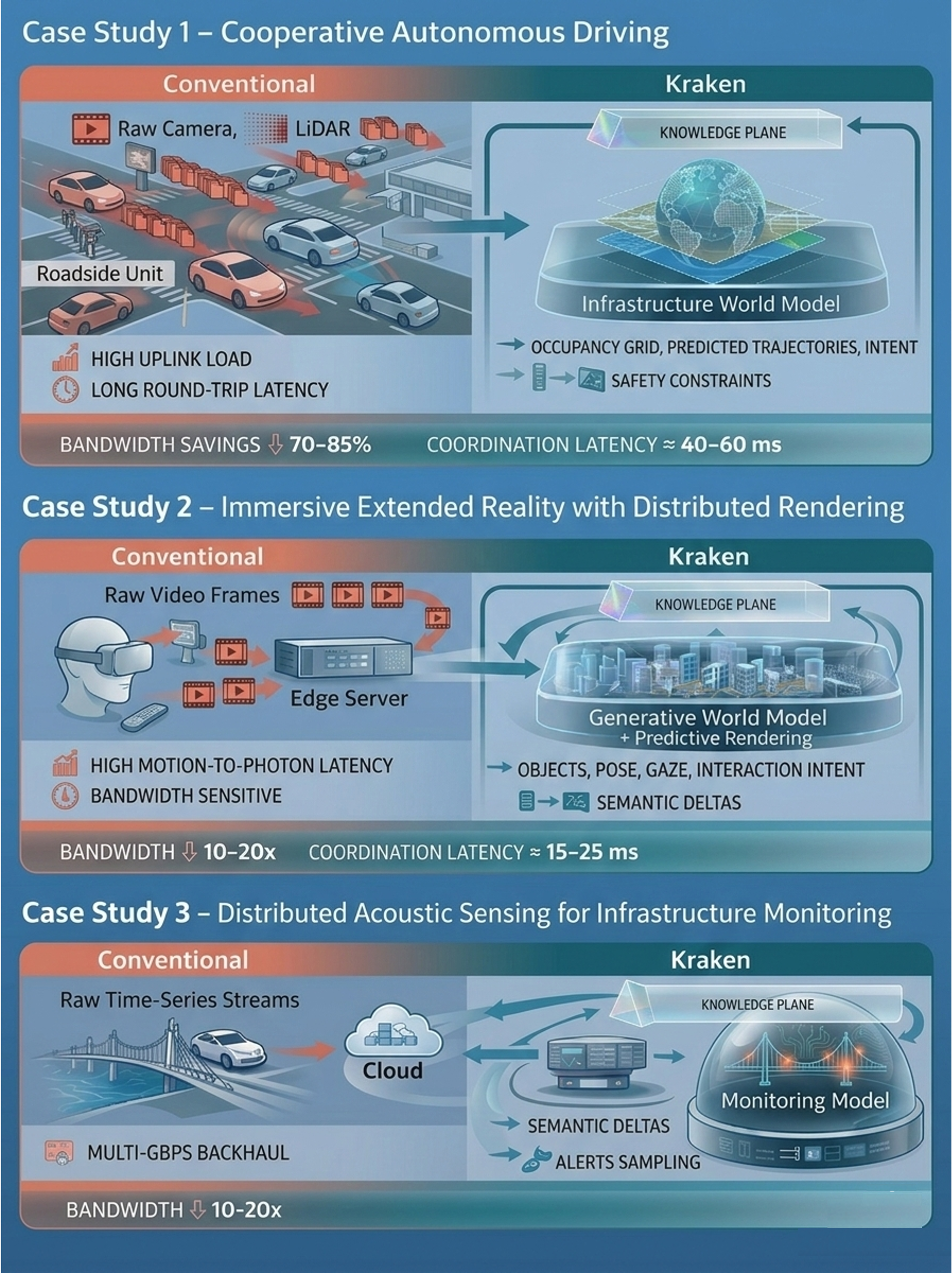}
    \caption{Three representative Kraken scenarios: autonomous driving with semantic coordination, immersive XR with predictive rendering, and infrastructure monitoring with edge intelligence. In each case, semantic abstraction occurs at the edge, generative reasoning operates at infrastructure nodes, and the Knowledge Plane maintains global consistency across distributed agents.}
    \label{fig:scenarios}
\end{figure}

To illustrate how the proposed architecture operates in practice, we consider three representative 6G scenarios. Each example shows how semantic communication, generative reasoning, and goal-oriented coordination together improve network efficiency, responsiveness, and reliability. The three scenarios and their interaction with the Kraken architecture are summarized in Fig.~\ref{fig:scenarios}.

\subsection{Scenario A: Autonomous Vehicles at a Blind Intersection}

\textit{The situation:}  
Four autonomous vehicles approach an intersection with obstructed visibility. To avoid collisions, they must determine the crossing order and adjust speeds within a few tens of milliseconds, requiring both extremely low latency and efficient information exchange.

\textit{Conventional approach:}  
In traditional architectures, each vehicle transmits raw sensor streams such as camera feeds and LiDAR point clouds to a nearby roadside unit or edge server. The infrastructure reconstructs the scene, computes coordination decisions, and returns instructions to the vehicles. This process may consume 50--100 Mbps of uplink bandwidth per vehicle and introduce round-trip delays of 150--200 ms, which are both bandwidth-intensive and too slow for reliable real-time coordination.

\textit{Kraken approach:}  
With Kraken, each vehicle hosts a Generative Network Agent that performs local perception and semantic abstraction. Instead of transmitting raw sensor streams, the vehicle generates compact knowledge objects such as occupancy grids, predicted trajectories with confidence levels, and maneuver intents. These descriptors, typically only a few kilobytes, are transmitted with semantic prioritization through the Infrastructure Plane. A roadside infrastructure agent aggregates the descriptors, uses its generative world model to detect potential trajectory conflicts, and broadcasts coordinated intent messages. Vehicles then adjust speed and crossing time accordingly, enabling distributed and proactive coordination.

\textit{What it enables:}  
Semantic abstraction significantly reduces communication overhead while preserving coordination accuracy. Bandwidth demand can decrease by 70--85\%, while coordination latency remains below 100 ms. Even under degraded channel conditions, vehicles retain sufficient semantic information to maintain safe behavior.

\subsection{Scenario B: Immersive XR for Factory Maintenance}

\textit{The situation:}  
A maintenance technician wearing an augmented reality headset performs repairs on complex industrial equipment. Digital overlays highlight components, display instructions, and visualize hidden structures. To ensure a stable user experience, motion-to-photon latency must remain below approximately 20 ms.

\textit{Conventional approach:}  
In conventional systems, the headset streams captured video frames to an edge server where scene understanding and rendering occur. The rendered frames are then encoded and transmitted back to the device. This tight feedback loop is highly sensitive to network fluctuations, and even small delays can introduce visible lag and user discomfort.

\textit{Kraken approach:}  
With Kraken, the headset performs local semantic scene extraction through its embedded agent. Instead of transmitting raw video frames, the device sends compact descriptors containing detected objects, user pose, gaze direction, and interaction intent. These descriptors are transmitted to an edge server hosting an infrastructure-level agent. Using a generative world model and cached environment maps, the edge reconstructs the scene and synthesizes rendered frames. Predictive rendering anticipates view changes from gaze trajectories and prepares frames in advance, reducing sensitivity to latency.

\textit{What it enables:}  
By transmitting semantic descriptors instead of full video streams, bandwidth consumption can decrease by an order of magnitude. In many cases, transmitted data is reduced by 10--20$\times$ while maintaining motion-to-photon latencies around 15--25 ms, resulting in a stable and responsive immersive experience.

\subsection{Scenario C: Bridge Health Monitoring}

\textit{The situation:}  
A city deploys distributed acoustic sensing along a suspension bridge to continuously monitor structural health. Fiber-optic sensors generate high-frequency time-series signals used to detect fatigue, abnormal vibrations, or structural damage.

\textit{Conventional approach:}  
Traditional monitoring systems transmit raw sensing streams to centralized cloud servers where signal processing and anomaly detection are performed. This approach creates heavy backhaul demands and may introduce delays that are too slow for timely detection of critical structural events.

\textit{Kraken approach:}  
Within the Kraken architecture, sensing nodes host lightweight Generative Network Agents that perform semantic abstraction locally. Instead of transmitting raw waveforms, sensors extract event descriptors including timestamps, anomaly scores, signal signatures, and confidence estimates. These compact knowledge objects are transmitted to infrastructure agents maintaining generative world models trained on historical structural responses. When deviations from normal behavior occur, the system generates immediate alerts together with probabilistic explanations. The Knowledge Plane further correlates observations across sensing locations to identify propagating vibration modes or emerging structural instabilities.

\textit{What it enables:}  
Edge-level semantic abstraction significantly reduces data transmission while enabling near-real-time anomaly detection. Compression ratios approaching 100:1 are achievable, allowing continuous monitoring without overwhelming backhaul networks while supporting predictive maintenance through continuously updated infrastructure digital twins.

These scenarios illustrate how Kraken transforms the network from a passive data transport substrate into an active coordination layer capable of understanding context, anticipating system dynamics, and aligning resource allocation with application-level objectives. By combining semantic communication, generative reasoning, and goal-oriented optimization, the architecture enables distributed collective intelligence across heterogeneous 6G services.

\begin{figure*}[h]	
    \centering
    \includegraphics[width=0.7\linewidth]{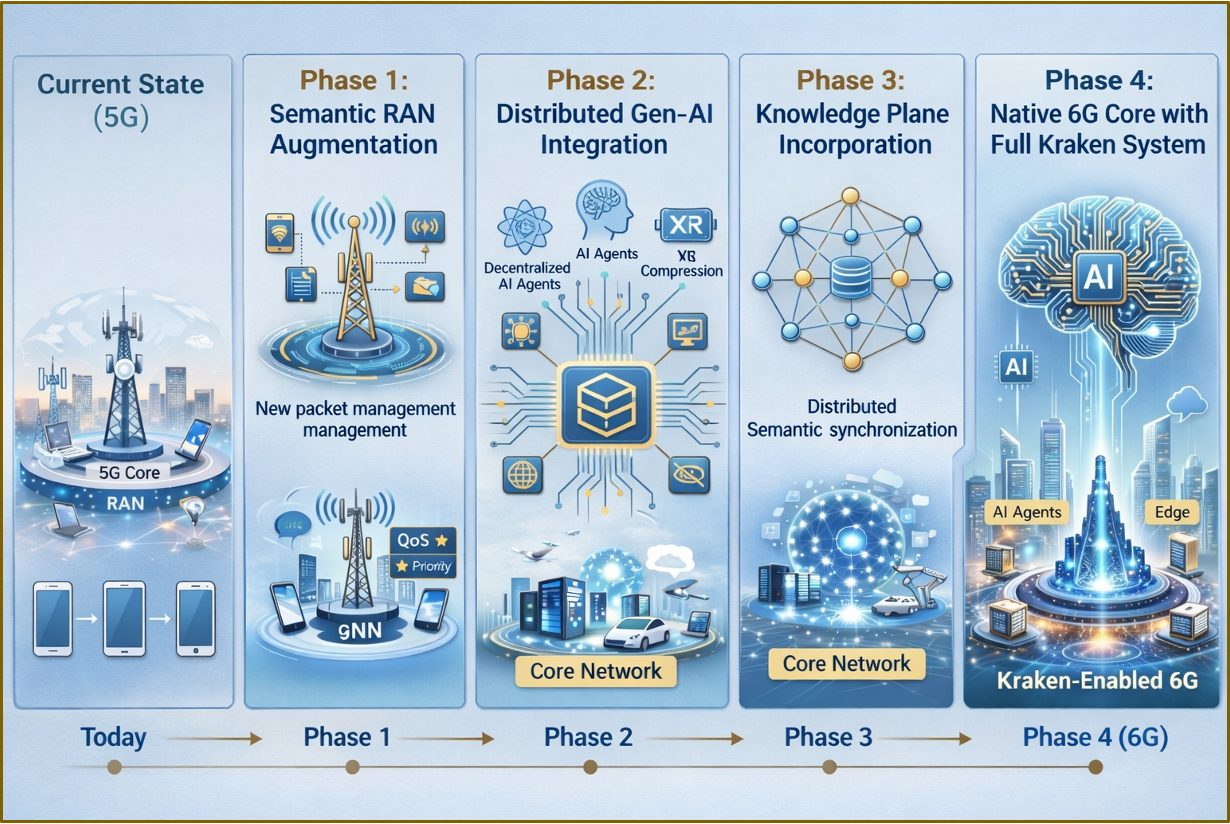}
\caption{Illustrative evolution from current 5G infrastructures toward Kraken-enabled 6G collective intelligence. The transition progresses through four stages: (1) semantic-aware enhancements in the RAN, (2) deployment of distributed generative agents at the edge, (3) emergence of a shared knowledge plane enabling coordinated intelligence, and (4) a native 6G architecture where semantic communication and collective reasoning become integral network capabilities.}
\label{evo}
\end{figure*}

\section{Evolution Toward Kraken-Enabled 6G}

The transition from current 5G infrastructures toward knowledge-centric 6G networks is unlikely to occur through an abrupt architectural shift. A more realistic path involves gradual evolution in which new capabilities are progressively integrated into existing systems. Introducing semantic awareness, generative reasoning, and knowledge-driven coordination requires alignment with deployed infrastructure and ongoing standardization efforts. Rather than redesigning networks from scratch, these capabilities can emerge incrementally while preserving compatibility with operational systems. Fig.~\ref{evo} illustrates a possible pathway through which intelligence progressively permeates the network stack.

\subsection{Stage 1: Semantic-Aware Enhancements}

The first stage introduces lightweight semantic awareness into existing 5G radio access networks. Applications provide indicators describing the relevance of transmitted information to the underlying task, complementing conventional QoS parameters and allowing network elements to prioritize transmissions according to their contribution to application outcomes. Even modest semantic annotations can improve scheduling efficiency, increase congestion robustness, and reduce the transmission of low-value data while maintaining compatibility with the existing 5G core architecture.

\subsection{Stage 2: Edge-Level Generative Intelligence}

The second stage introduces distributed intelligence through generative agents operating at the network edge. Initially deployed in observational or advisory roles, these agents analyze network telemetry to learn traffic dynamics, mobility patterns, and service behavior. As models mature, they begin supporting network functions through predictive analytics, adaptive resource allocation, and semantic compression, gradually demonstrating the benefits of predictive coordination alongside existing control mechanisms.

\subsection{Stage 3: Emergence of a Knowledge Plane}

As edge intelligence matures, coordination among agents becomes increasingly important. This stage introduces a shared knowledge layer enabling agents to exchange structured representations rather than isolated measurements. Knowledge graphs, shared semantic embeddings, and intent descriptions provide a common context that aligns local decisions with system-level objectives, enabling coordinated optimization across communication, sensing, and computing domains.

\subsection{Stage 4: Native Knowledge-Centric 6G Networks}

In the final stage, knowledge-centric operation becomes a native property of the network architecture. Semantic representations, generative reasoning, and goal-oriented coordination are integrated directly into communication and control mechanisms. Network entities, from devices to infrastructure nodes, participate in distributed cognitive processes combining perception, reasoning, and action, allowing the network to actively support collective intelligence across connected systems.

This evolutionary pathway enables the gradual introduction of new capabilities while preserving compatibility with existing deployments. Each stage provides immediate operational benefits while preparing the foundation for more advanced forms of distributed intelligence in future 6G networks.

\section{The Road Ahead: Five Grand Challenges}

The Kraken architecture illustrates how knowledge-centric networking could operate in future 6G systems. Transforming this vision into large-scale operational reality requires addressing several fundamental research challenges. Moving from bit-centric communication toward knowledge-driven coordination raises open questions spanning theory, system design, hardware realization, security, and evaluation. Progress in these areas will require close collaboration across the communications, networking, and artificial intelligence communities.

\subsection{Semantic Information Theory}

Classical information theory established the limits of reliable bit transmission over noisy channels but does not explicitly capture task relevance or semantic meaning. In knowledge-centric networks, the objective shifts from perfect symbol reconstruction to preserving information necessary for successful task execution. Developing a theory of semantic information therefore remains a key challenge, particularly in understanding the trade-offs between semantic compression, communication cost, and task-level performance in distributed systems.

\subsection{Multi-Agent Goal Alignment}

Kraken relies on distributed Generative Network Agents that coordinate through semantic information exchange rather than centralized optimization. While this improves scalability and responsiveness, independently operating agents must still remain aligned with shared system objectives. Ensuring stable and coherent behavior under uncertainty, partial observability, and asynchronous communication therefore remains an open problem, requiring mechanisms for distributed goal alignment, stability, and conflict resolution in large multi-agent environments.

\subsection{Hardware-Aware Generative Intelligence}

Generative world models and transformer-based architectures offer powerful reasoning capabilities but often exceed the computational limits of edge devices and communication infrastructure. Deploying such models in latency-critical environments requires careful hardware–software co-design. Future work must therefore develop lightweight generative models that maintain reasoning capability while meeting strict latency, energy, and memory constraints, potentially through model compression, sparse attention mechanisms, and specialized AI accelerators.

\begin{figure}[h]
    \centering
    \includegraphics[width=0.9\linewidth]{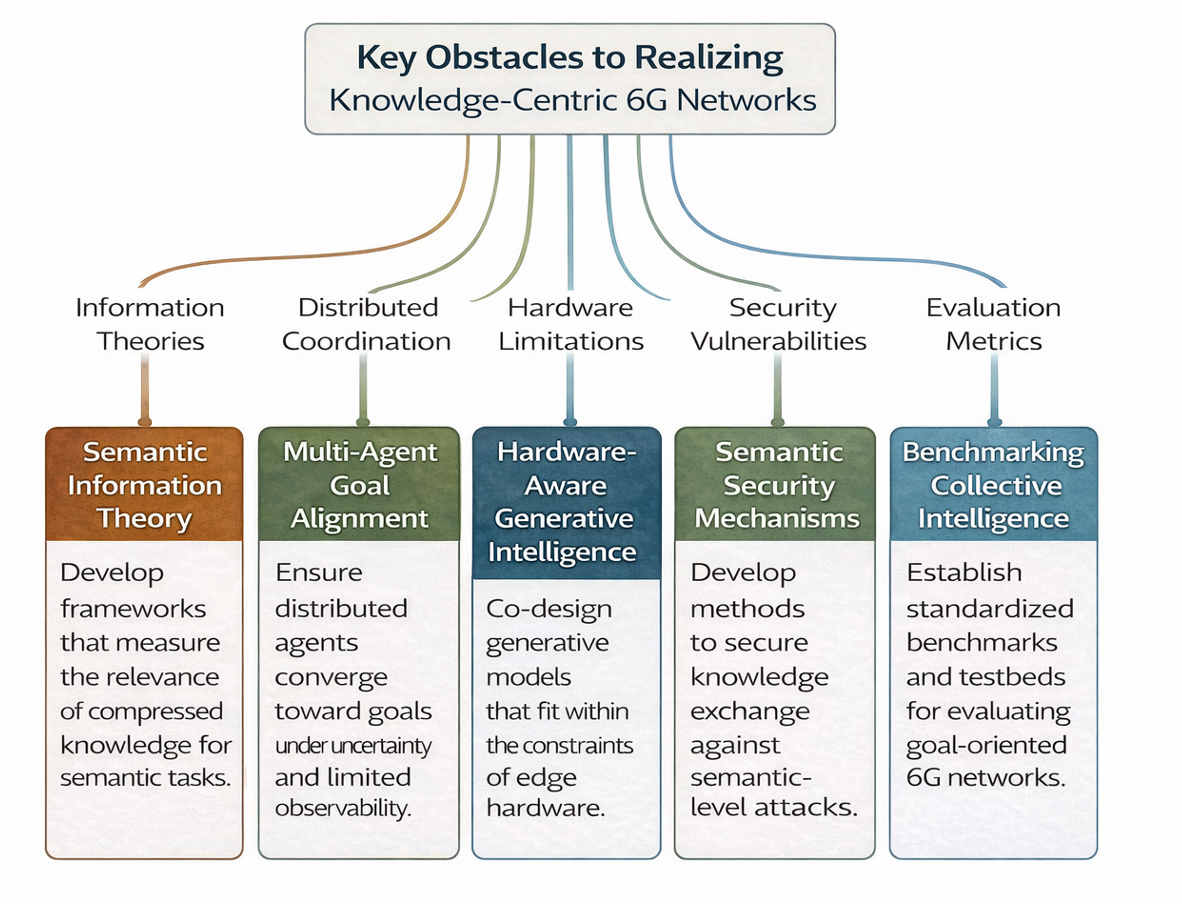}
\caption{Grand challenges on the path toward knowledge-centric 6G networks. 
Future systems must address fundamental questions in semantic information theory, 
multi-agent goal alignment, hardware-efficient generative intelligence, 
semantic security mechanisms, and benchmarking methodologies capable of 
evaluating collective intelligence in large-scale network environments.}
\label{fig5}
\end{figure}

\subsection{Semantic Security Mechanisms}

Knowledge-centric communication introduces security challenges beyond traditional packet-level threats. When networks exchange semantic representations rather than raw data, adversaries may attempt to inject misleading knowledge objects, manipulate confidence values, or disrupt reasoning processes across distributed agents. Protecting such systems therefore requires security mechanisms operating directly in semantic space, including provenance verification of knowledge objects, cross-agent validation, and anomaly-detection techniques capable of identifying semantic manipulation.

\subsection{Benchmarking Collective Intelligence}

Evaluating knowledge-centric network architectures also requires new benchmarking methodologies. Conventional metrics such as throughput and delay capture only part of system behavior and cannot fully reflect the effectiveness of semantic coordination or goal-oriented optimization. Future evaluation frameworks should therefore incorporate metrics such as semantic efficiency, task success probability, coordination latency, and resource overhead, supported by standardized datasets, experimental testbeds, and shared benchmarks for collective intelligence in large-scale networks.
\section{Conclusion}
Kraken provides a coherent architectural direction for the transition from today’s data-centric networks toward knowledge-centric collective intelligence. By integrating semantic communication, generative reasoning, and goal-oriented optimization within a three-plane framework, it transforms passive infrastructure into cognitive systems capable of anticipation, adaptation, and intent alignment. The evolution from 5G to Kraken-enabled 6G is gradual rather than disruptive. Each stage introduces semantic awareness, distributed generative reasoning, and intent-driven coordination while delivering immediate operational value. Early examples in autonomous mobility, immersive XR, and sensing infrastructures illustrate how these principles reduce communication overhead while improving responsiveness and system-level efficiency. Realizing this vision will require interdisciplinary advances in semantic information theory, hardware-aware generative models, secure semantic communication, and evaluation frameworks for collective intelligence. Kraken provides the architectural compass for this transition toward knowledge-centric 6G networks.

	\bibliographystyle{IEEEtran}
	\bibliography{ref}

\begin{thebibliography}{10}
\providecommand{\url}[1]{#1}
\csname url@samestyle\endcsname
\providecommand{\newblock}{\relax}
\providecommand{\bibinfo}[2]{#2}
\providecommand{\BIBentrySTDinterwordspacing}{\spaceskip=0pt\relax}
\providecommand{\BIBentryALTinterwordstretchfactor}{4}
\providecommand{\BIBentryALTinterwordspacing}{\spaceskip=\fontdimen2\font plus
\BIBentryALTinterwordstretchfactor\fontdimen3\font minus
  \fontdimen4\font\relax}
\providecommand{\BIBforeignlanguage}[2]{{%
\expandafter\ifx\csname l@#1\endcsname\relax
\typeout{** WARNING: IEEEtran.bst: No hyphenation pattern has been}%
\typeout{** loaded for the language `#1'. Using the pattern for}%
\typeout{** the default language instead.}%
\else
\language=\csname l@#1\endcsname
\fi
#2}}
\providecommand{\BIBdecl}{\relax}
\BIBdecl

\bibitem{Akyildiz2022_XR}
\BIBentryALTinterwordspacing
I.~F. Akyildiz and H.~Guo, ``Wireless communication research challenges for
  extended reality (xr),'' \emph{ITU Journal on Future and Evolving
  Technologies}, vol.~3, no.~2, pp. 273--287, April 2022. [Online]. Available:
  \url{https://doi.org/10.52953/QGKV1321}
\BIBentrySTDinterwordspacing

\bibitem{9955525}
D.~Gündüz, Z.~Qin, I.~E. Aguerri, H.~S. Dhillon, Z.~Yang, A.~Yener, K.~K.
  Wong, and C.-B. Chae, ``Beyond transmitting bits: Context, semantics, and
  task-oriented communications,'' \emph{IEEE Journal on Selected Areas in
  Communications}, vol.~41, no.~1, pp. 5--41, 2023.

\bibitem{10597087}
E.~C. Strinati, P.~Di~Lorenzo, V.~Sciancalepore, A.~Aijaz, M.~Kountouris,
  D.~Gündüz, P.~Popovski, M.~Sana, P.~A. Stavrou, B.~Soret, N.~Cordeschi,
  S.~Scardapane, M.~Merluzzi, L.~Zanzi, M.~B. Renato, T.~Quek, N.~D. Pietro,
  O.~Forceville, F.~Costanzo, and P.~Li, ``Goal-oriented and semantic
  communication in 6g ai-native networks: The 6g-goals approach,'' in
  \emph{2024 Joint European Conference on Networks and Communications \& 6G
  Summit (EuCNC-6G Summit)}, 2024, pp. 1--6.

\bibitem{akyildiz2026kraken}
I.~F. Akyildiz and T.~Bilen, ``Kraken: Architecting generative, semantic, and
  goal-oriented network management for 6g wireless systems,'' \emph{submitted
  to Proceedings of the IEEE}, March 2026.

\bibitem{8723589}
E.~Bourtsoulatze, D.~Burth~Kurka, and D.~Gündüz, ``Deep joint source-channel
  coding for wireless image transmission,'' \emph{IEEE Transactions on
  Cognitive Communications and Networking}, vol.~5, no.~3, pp. 567--579, 2019.

\bibitem{8714026}
N.~C. Luong, D.~T. Hoang, S.~Gong, D.~Niyato, P.~Wang, Y.-C. Liang, and D.~I.
  Kim, ``Applications of deep reinforcement learning in communications and
  networking: A survey,'' \emph{IEEE Communications Surveys \& Tutorials},
  vol.~21, no.~4, pp. 3133--3174, 2019.

\bibitem{10867389}
R.~S and S.~E. Selvi, ``Anomaly detection in network traffic data using deep
  learning,'' in \emph{2024 2nd International Conference on Computing and Data
  Analytics (ICCDA)}, 2024, pp. 1--5.

\bibitem{10644029}
H.~Zhou, Y.~Deng, X.~Liu, N.~Pappas, and A.~Nallanathan, ``Goal-oriented
  semantic communications for 6g networks,'' \emph{IEEE Internet of Things
  Magazine}, vol.~7, no.~5, pp. 104--110, 2024.

\bibitem{Oliehoek2016}
F.~A. Oliehoek and C.~Amato, \emph{A Concise Introduction to Decentralized
  POMDPs}, ser. SpringerBriefs in Intelligent Systems.\hskip 1em plus 0.5em
  minus 0.4em\relax Springer Publishing Company, Incorporated, June 2016.

\bibitem{doi:https://doi.org/10.1002/9781119847083.ch4}
\BIBentryALTinterwordspacing
\emph{O-RAN Alliance Architecture}.\hskip 1em plus 0.5em minus 0.4em\relax John
  Wiley \& Sons, Ltd, 2024, ch.~4, pp. 59--86. [Online]. Available:
  \url{https://onlinelibrary.wiley.com/doi/abs/10.1002/9781119847083.ch4}
\BIBentrySTDinterwordspacing

\bibitem{WANG2024124679}
\BIBentryALTinterwordspacing
T.~Wang, G.~Qi, and T.~Wu, ``Kgroot: A knowledge graph-enhanced method for root
  cause analysis,'' \emph{Expert Systems with Applications}, vol. 255, p.
  124679, 2024. [Online]. Available:
  \url{https://www.sciencedirect.com/science/article/pii/S095741742401546X}
\BIBentrySTDinterwordspacing

\bibitem{10466747}
Y.~Chen, R.~Li, Z.~Zhao, C.~Peng, J.~Wu, E.~Hossain, and H.~Zhang, ``Netgpt: An
  ai-native network architecture for provisioning beyond personalized
  generative services,'' \emph{IEEE Network}, vol.~38, no.~6, pp. 404--413,
  2024.

\bibitem{3gpp22_261}
{3GPP}, ``Service requirements for the 5g system,'' 3rd Generation Partnership
  Project (3GPP), Tech. Rep. TS 22.261, 2021.

\bibitem{3gpp23_288}
{3GPP{}}, ``Architecture enhancements for 5g system (5gs) to support network
  data analytics services,'' 3rd Generation Partnership Project (3GPP), Tech.
  Rep. TS 23.288, 2023.

\bibitem{polese2023understanding}
M.~Polese, L.~Bonati, S.~D’oro, S.~Basagni, and T.~Melodia, ``Understanding
  o-ran: Architecture, interfaces, algorithms, security, and research
  challenges,'' \emph{IEEE Communications Surveys \& Tutorials}, vol.~25,
  no.~2, pp. 1376--1411, 2023.

\end{thebibliography}

\begin{IEEEbiography}[{\includegraphics[width=1in,height=1.3in,clip,keepaspectratio]{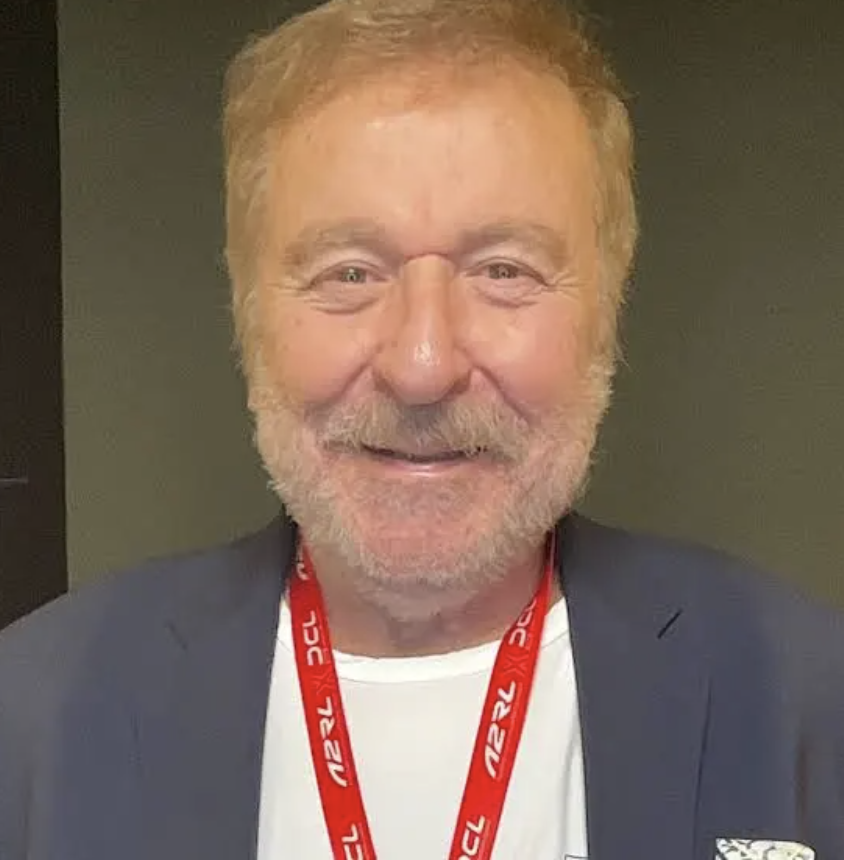}}]{Ian F. Akyildiz}
(Life Fellow, IEEE) received his B.S., M.S., and Ph.D. degrees in electrical and computer engineering from the University of Erlangen–Nürnberg, Germany, in 1978, 1981, and 1984, respectively. From 1985 to 2020, he held the Ken Byers Chair Professorship at Georgia Tech, where he directed the Broadband Wireless Networking Laboratory. A visionary entrepreneur, he is the President of Truva Inc. and a key advisor to global institutions like TII (Abu Dhabi) and Odine Labs (Istanbul). Since 2020, he has served as the founding Editor-in-Chief of the ITU Journal on Future and Evolving Technologies. His pioneering research spans 6G/7G systems, molecular communication, terahertz technology, and underwater networking. As of March 2026, he holds an H-index of 146 with over 155,000 citations. Dr. Akyildiz is an ACM Fellow and recipient of prestigious honors, including the Humboldt (Germany) and TÜBİTAK (Türkiye) Awards.
\end{IEEEbiography}

\begin{IEEEbiography}[{\includegraphics[width=1in,height=1.3in,clip,keepaspectratio]{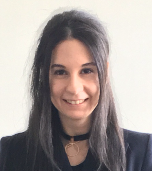}}]{Tuğçe Bilen}
(Member, IEEE) received her B.Sc., M.Sc., and Ph.D. degrees in Computer Engineering from Istanbul Technical University (ITU) in 2015, 2017, and 2022, respectively. She is currently an Assistant Professor in the Department of Artificial Intelligence and Data Engineering at ITU, where she previously served as a Research and Teaching Assistant. Her doctoral research earned several prestigious honors, including the 2025 IEEE Turkey Section Ph.D. Thesis Award, the 2023 Turkish Academy of Sciences (TÜBA) First Prize in Science and Technology, the 2023 Serhat Özyar Young Scientist Honorary Award, and the 2022 ITU Best Ph.D. Thesis Award. Her research focuses on 6G networks, Knowledge-Defined Networking (KDN), AI-driven network management, and digital twins, specifically integrating intelligent systems into future architectures. Dr. Bilen also serves as a reviewer for leading international journals.
\end{IEEEbiography}

\end{document}